\def\ra{\rangle}

\def\be{\begin{equation}}
\def\ee{\end{equation}}
\def\ba{\begin{array}}
\def\ea{\end{array}}

\def\Rb{{I\!\! R}}

\def\Cb{{\Bbb C}}

\documentstyle[12pt]{article}
\begin{document}
\input amssym.def
\setcounter{page}{1}
\centerline{\large\bf Matrix Tensor Product Approach to the Equivalence of}
\vspace{2ex}
\centerline{\large\bf Multipartite States under Local Unitary Transformations} \vspace{4ex}
\begin{center}

X. H. Gao$^{1}$, ~S. Alberverio$^{2}$,  ~S. M. Fei$^{1,2,3}$, ~Z.X. Wang$^{1}$

\vspace{2ex}
\begin{minipage}{5in}

{\small $~^{1}$ Department of Mathematics, Capital Normal
University, Beijing 100037}

{\small $~^{2}$ Institut f\"ur Angewandte Mathematik,
Universit\"at Bonn, D-53115}

{\small $~^{3}$ Max-Planck-Institute for Mathematics in the
Sciences, 04103 Leipzig}

\end{minipage}
\end{center}

 \vskip 1 true cm
\parindent=18pt
\parskip=6pt
\begin{center}
\begin{minipage}{4.5in}
\vspace{3ex} \centerline{\large Abstract} \vspace{4ex}

The equivalence of multipartite quantum mixed states under
local unitary transformations is studied. A criterion for the
equivalence of non-degenerate mixed multipartite quantum states under
local unitary transformations is presented.

\bigskip
\bigskip

\vfill
\smallskip

\end{minipage}
\end{center}

\bigskip
\medskip
\bigskip
\medskip

Quantum entangled states are playing fundmental roles
in quantum information processing such as quantum computation,
quantum teleportation, dense coding, quantum cryptographic schemes£¬
quantum error correction, entanglement swapping, and remote state
preparation (RSP) etc.. However the theory of quantum entanglement
is still far from being satisfied.
To quantify the degree of entanglement a
number of entanglement measures have been proposed for bipartite states.
Most of these proposed measures of entanglement
involve extremizations which are difficult to handle analytically.
For multipartite case how to give a well defined
measure is still under discussion.
For general mixed states till now we don't even have an operational
criterion to verify whether a state is separable or not.
As a matter of fact, the degree of entanglement
of a multipartite quantum system remains invariant under local
unitary transformations of every subsystems. Therefore the quantum
states can be classified according to the local unitary
transformations. Nevertheless an explicit picture of the orbits
(geometry and topology) under such transformations is not yet
ready. We even don't have a general (operational) criterion to
verify if two mixed states are equivalent or not under local
unitary transformations.

One approach in dealing
with the equivalence of quantum states under local unitary transformations
is to find the complete set of invariants under
local unitary transformations. Two states are equivalent under
local unitary transformations if and only if they have the same
values of all these invariants. The method developed in
\cite{Rains,Grassl}, in principle, allows one to compute all the
invariants of local unitary transformations, though in general it
is not operational. The invariants for general
two-qubit and three qubits systems have been studied in \cite{makhlin,linden}.
In \cite{generic} a complete set of invariants is
presented for bipartite generic mixed states. In \cite{goswami} a
complete set of invariants under local unitary transformations is
presented for rank-2 and multiplicity free mixed (bipartite) states. The invariants
for pure tripartite states are also studied in \cite{tri1,tri2}.

In \cite{jing-eq} another approach is presented in investigating
the equivalence problem in terms of
fixed point subgroup and tensor decomposability of certain
matrices. The problem is reduced to verify whether a certain matrix
is rank one or not. A criterion for the equivalence of two non-degenerate
mixed bipartite states under local unitary
transformations has been presented.

It is rather difficult to deal with the equivalence problem
of multipartite states in terms of invariants approach.
The number of invariants increases quickly when the
subsystems increase. Nevertheless, we find that the approach in
\cite{jing-eq} can be easily generalized to multipartite case.
In this note we investigate the equivalence of multipartite states
under local unitary transformations by using the method in
\cite{jing-eq}. An operational criterion for the equivalence of non-degenerate
mixed multipartite states under local unitary
transformations is presented.

We first consider tripartite case.
Let $H_1$ (resp. $H_2$ and $H_3$) be an $M$ (resp. $N$ and
$P$)-dimensional complex Hilbert space, with $\vert e_i\rangle$,
$i=1,...,M$ (resp. $\vert f_j\rangle$, $j=1,...,N$ and $\vert
g_k\rangle$, $k=1,...,P$), as an orthonormal basis. A general pure
state on $H_1\otimes H_2\otimes H_3$ is of the form
\begin{equation}\label{mmm}
\vert\Psi\rangle=\sum_{i=1}^M\sum_{j=1}^N\sum_{k=1}^P a_{ijk}\vert
e_i\rangle \otimes \vert f_j\rangle\otimes \vert
g_k\rangle,~~~~~~a_{ijk}\in\Cb
\end{equation}
with the normalization
$\displaystyle\sum_{i=1}^M\sum_{j=1}^N\sum_{k=1}^P
a_{ijk}a_{ijk}^\ast=1$ ($\ast$ denoting complex conjugation).
A tripartite quantum mixed state on $H_1\otimes H_2\otimes H_3$ is
described by a density matrix $\rho$ which can be decomposed
according to its eigenvalues and eigenvectors:
$
\rho=\sum_{i=1}^{MNP}\lambda_i\vert\nu_i\rangle\langle\nu_i\vert,
$
where $\lambda_i$ are the eigenvalues and $\vert\nu_i\rangle$,
$i=1,...,MNP$, the corresponding eigenvectors of the form
(\ref{mmm}).

Two density matrices $\rho$ and $\rho^\prime$ are said to be
equivalent under local unitary transformations if there exist
unitary operators $U_1$ on $H_1$, $U_2$ on $H_2$ and $U_3$ on
$H_3$ such that
\be
\rho^\prime=(U_1\otimes U_2\otimes
U_3)\rho(U_1\otimes U_2\otimes U_3)^\dag.
\ee

For a Hermitian matrix $A$ on $H_1\otimes H_2\otimes H_3$, the set
of commuting matrices $B$ such that $AB=BA$ is called the
commutant of $A$, denoted as $C(A)$. Obviously $C(A)$
is a subalgebra. We call the set of unitary
matrices $U$ such that $UA=AU$ the fixed point
subgroup of $A$, denoted as $C_U(A)$, which is a subgroup of the
unitary group of all unitary matrices.
In the following we say that a matrix $V$ on $H_1\otimes H_2\otimes H_3$
is tensor decomposable if it can be written as $V=V_1\otimes V_2\otimes V_3$ for $V_1\in
End(H_1)$, $V_2\in End(H_2)$, $V_3\in End(H_3)$.

If two density matrices $\rho$ and $\rho^\prime$ are equivalent
under local unitary transformations, they must have the same set
of eigenvalues $\lambda_i$, $i=1,...,n$. Let $X$ and $Y$ be the
unitary matrices that diagonalize $\rho$ and $\rho^\prime$
respectively, \be\label{xy} \rho=X\Lambda
X^\dag,~~~~~\rho^\prime=Y\Lambda Y^\dag, \ee where
$\Lambda=diag(\lambda_1,\lambda_2,...,\lambda_{MNP})$.

{\sf [Lemma 1].} Let $G$ be the fixed point unitary subgroup
associated with $\rho$. Then $\rho^\prime$ is equivalent to $\rho$
under local unitary transformations if and only if the coset
$GXY^\dag$ contains a unitary tensor decomposable matrix.

{\sf[Proof].} Suppose $\rho^\prime=(U_1\otimes U_2\otimes U_3)\rho(U_1\otimes U_2\otimes U_3)^\dag$, then
$\rho^\prime=Y\Lambda Y^\dag=Y X^\dag \rho X Y^\dag$. Hence
$Y X^\dag \rho X Y^\dag=(U_1\otimes U_2\otimes U_3)\rho(U_1\otimes U_2\otimes U_3)^\dag$, or
$$
(U_1\otimes U_2\otimes U_3)^\dag Y X^\dag \rho= \rho(U_1\otimes U_2\otimes U_3)^\dag Y X^\dag.
$$
That is, $(U_1\otimes U_2\otimes U_3)^\dag Y X^\dag\in C_U(\rho)$, and
$C_U(\rho)X Y^\dag=YX^\dag C_U(\rho)$ contains a unitary tensor decomposable
element $U_1\otimes U_2\otimes U_3$.

Conversely, assume $GXY^\dag$ contains a tensor decomposable
element $U_1\otimes U_2\otimes U_3$. We have then $UXY^\dag=U_1\otimes U_2\otimes U_3$,
$U\rho=\rho U$ and $U$ is unitary. Therefore
$$\ba{rcl}
\rho^\prime&=&Y\Lambda Y^\dag=(U_1\otimes U_2\otimes U_3)^\dag U X
\Lambda X^\dag U^\dag (U_1\otimes U_2\otimes U_3)\\
&=&(U_1\otimes U_2\otimes U_3)^\dag U \rho U^\dag (U_1\otimes U_2\otimes U_3)
=(U_1\otimes U_2\otimes U_3)^\dag \rho (U_1\otimes U_2\otimes U_3).
\ea
$$
\hfill $\Box$

Let $Z$ be an a matrix on $H_1\otimes H_2\otimes H_3$. If we view
$Z$ as a matrix on spaces $H_1$ and $H_2\otimes H_3$, it is an
$M\times M$ block matrix with each block of size
$NP\times NP$. Its realigned matrix $\tilde{Z}_{1|23}$ is defined
by
$$
\tilde{Z}_{1|23}=[vec(Z_{11}),\cdots,vec(Z_{1M}),\cdots,
vec(Z_{M1}),\cdots, vec(Z_{MM})]^t,
$$
Taking $Z$ as a matrix on spaces $H_1\otimes H_2$ and $H_3$, we have that
$Z$ is an $MN\times MN$ block matrix with each block of size
$P\times P$, and the realigned matrix $\tilde{Z}_{12|3}$ is of the form
$$
\tilde{Z}_{12|3}=[vec(Z_{11}),\cdots,vec(Z_{1MN}),\cdots,
vec(Z_{MN1}),\cdots, vec(Z_{MNMN})]^t,
$$
where for any $M\times N$ matrix $A$  with  entries $a_{ij}$,
$vec(A)$ is defined to be
$$
vec(A)=[a_{11},\cdots,a_{1N},a_{21},\cdots,a_{2N},\cdots,a_{M1},\cdots,
a_{MN}]^t.
$$
The above operation of realignment could be defined in an alternative way,
$$
(\tilde{Z}_{1|23})_{im,jknp}=(Z)_{ijk,mnp}\,,~~~~~
(\tilde{Z}_{12|3})_{ijmn,kp}=(Z)_{ijk,mnp}.
$$
It is shown that a matrix $V$ can be expressed as the tensor
product of two matrices $V_1$ and $V_2$, $V=V_1\otimes V_2$, if and
only if \cite{kropro}
\be \label{ch}
\tilde{V}=vec(V_1)vec(V_2)^t .
\ee
Moreover \cite{tri1}, for an
$MN\times MN$ unitary matrix $U$, if $U$ is a unitarily
decomposable matrix, then the rank of $\tilde{U}$ is one,
$r(\tilde{U})=1$. Conversely if $r(\tilde{U})=1$, there exists an
$M\times M$ matrix $U_1$ and an $N\times N$ matrix $U_2$, such
that $U=U_1\otimes U_2$ and \be\label{kk}
U_1U_1^{\dag}=U_1^{\dag}U_1=k^{-1}I_M,~~~~U_2U_2^{\dag}=U_2^{\dag}U_2=kI_N,
\ee where $I_N$ (resp. $I_M$) denotes the $N\times N$ (resp.
$M\times M$) identity matrix, $k>0$, and $U$ is a unitary tensor
decomposable matrix.

{\sf [Lemma 2].} For an $MNP\times MNP$ unitary matrix $U$
on $H_1\otimes H_2\otimes H_3$, $U$
is a unitary decomposable matrix if and only if the ranks of
$\tilde{U}_{1|23}, \tilde{U}_{12|3}$ are one, i.e.,
$r(\tilde{U}_{1|23})=r(\tilde{U}_{12|3})=1.$

{\sf [Proof].} Suppose $U=U_1\otimes U_2\otimes U_3$, then
$\tilde{U}_{1|23}=vec(U_1)vec(U_2\otimes U_3)^t$,
$\tilde{U}_{12|3}=vec(U_1\otimes U_2)vec(U_3)^t$, obviously
$r(\tilde{U}_{1|23})=r(\tilde{U}_{12|3})=1.$

Conversely, if $r(\tilde{U}_{1|23})=1$, there exists an
$M\times M$ matrix $U_1$ and an $NP\times NP$ matrix $U_{23}$,
such that $U=U_1\otimes U_{23}$. Therefore
\be\label{x} \tilde{U}_{12|3}=\left( \ba{c}
u_{11}\tilde{U}_{23}\\
u_{12}\tilde{U}_{23}\\
\vdots\\
u_{MM}\tilde{U}_{23} \ea \right),
\ee
where $u_{ij} í,j=1, \cdots, M$ are the entries of $U_1$.
$r(\tilde{U}_{12|3})=1$ if and only if $r(\tilde{U}_{23})=1$,
that is, there exists an $N\times N$
matrix $U_2$ and a $P\times P$ matrix $U_3$, such that
$U_{23}=U_2\otimes U_3$, and $U$ is a unitary tensor decomposable
matrix. \hfill $\Box$

Let $\rho$ and $\rho^\prime$ be two density matrices with
orthonormal unitary matrices $X$ and $Y$ as given in (\ref{xy}).
Set
\be\label{v0}
V_0=X\left( \ba{cccc}
A_{n_1}&0&\cdots&0\\
0&A_{n_2}&\cdots&0\\
\vdots&&\ddots&\vdots\\
0&\cdots&\cdots&A_{n_r} \ea \right)Y^\dag,
\ee
where $n_i$, $i=1,2,...,r$, stands for the geometric multiplicity
of the eigenvalue $\lambda_i$ of $\rho$, $\sum_1^r n_r=MNP$,
$A_{n_i}$ are some unitary $n_i\times n_i$ complex matrices. The conclusion
on bipartite case \cite{jing-eq} is still valid, namely,
$\rho$ and $\rho^\prime$ are equivalent under local unitary
transformations if and only if rank $r(\tilde{V_0})=1$ for
some unitary matrices $A_{n_i}$.
For the case that $\rho$ has distinct eigenvalues, we have

{\sf [Theorem 1].} If $\rho$ has distinct eigenvalues,
$\rho$ is equivalent to $\rho^\prime$ under local unitary transformations if and
only if
\be\label{v}
V=XDY^\dag,
\ee
$D=diag(e^{i\theta_1},\,e^{i\theta_2},...,e^{i\theta_{MNP}})$,
contains a unitary tensor decomposable element for some
$\theta_i\in \Rb$.

{\sf [Proof].} Let $X$ and $Y$ be the unitary matrices that
diagonalize $\rho$ and $\rho^\prime$ respectively,
$\rho=X\Lambda X^\dag,~\rho^\prime=Y\Lambda Y^\dag$, where
$\Lambda=diag(\lambda_1,\lambda_2,...,\lambda_{MNP})$.

As $\rho$ (resp. $\rho^\prime$) has distinct eigenvalues,
any set $X_1$ (resp. $Y_1$) of unitary eigenvectors corresponding
to the eigenvalues of $\rho$ (resp. $\rho^\prime$) can then be
obtained through the following equation: $X_1=XU$ (resp. $Y_1=YV$),
where the unitary matrix $U$ (resp. $V$) has the form:
$U=diag(e^{i\alpha_1},\,e^{i\alpha_2},...,e^{i\alpha_{MNP}})$
(resp.
$V=diag(e^{i\beta_1},\,e^{i\beta_2},...,e^{i\beta_{MNP}})$),
then $X_1D_1Y_1^\dag=XUD_1 V^\dag Y^\dag=XD Y^\dag$,
where $D_1=diag(e^{i\theta_1'},\,e^{i\theta_2'},...,e^{i\theta_{MNP}'})$,
$D=diag(e^{i\theta_1},\,e^{i\theta_2},...,e^{i\theta_{MNP}})$,
$\theta_i=\alpha_i+\theta_i'-\beta_i, i=1,2,...,MNP$,
hence $\rho$ is locally unitary equivalent to $\rho^\prime$ if and
only if $XDY^\dag$,
$D=diag(e^{i\theta_1},\,e^{i\theta_2},...,e^{i\theta_{MNP}})$,
contains a unitary tensor decomposable element for some
$\theta_i\in \Rb$. \hfill$\Box$

As an example let us consider a density matrix on $2\times 2\times 2$,
\be\label{exam1} \rho = \left(\ba{cccccccc}
1& 0& 0& 0& 0& 0& 0&-1\\
0& 1/a& 0& 0& 0& 0& 0& 0\\
0& 0& 1/b& 0& 0& 0& 0& 0\\
0& 0& 0& 1/c& 0& 0&0& 0\\
0& 0& 0& 0& c& 0& 0& 0\\
0& 0& 0&0& 0& b& 0& 0\\
0& 0& 0& 0& 0& 0& a& 0\\
-1& 0& 0& 0& 0&0& 0& 1
\ea\right),
\ee
\be\label{exam11} \rho^\prime = \left(\ba{cccccccc}
1& 0& 0& 0&
0&0& 0& 1\\0& a& 0& 0& 0& 0& 0& 0\\0& 0& b& 0& 0& 0&
       0& 0\\0& 0& 0& c& 0& 0& 0& 0\\0& 0& 0& 0& 1/c& 0& 0& 0\\0& 0& 0&
      0& 0& 1/b& 0& 0\\0& 0& 0& 0& 0& 0& 1/a& 0\\1& 0& 0& 0& 0& 0& 0& 1\ea\right),
\ee
$\rho^\prime$ is in fact a PPT entangled edge state \cite{prl}.
$\rho^\prime$ and $\rho$ have the same eigenvalue set.
They are not degenerate in the case $a\neq b\neq c\neq 1$
or $2$ or $1/2$. Calculating the unitary matrices $X$ and $Y$
that diagonalize $\rho$ and $\rho^\prime$, we have $\rho=X\Lambda X^\dag$,
$\rho^\prime=Y\Lambda Y^\dag$, where $\Lambda=diag(2,0,1/a,a,1/b,b,1/c,c)$. Denote
$D=diag(d_1,d_2,d_3,d_4,d_5,d_6,d_7,d_8)$, we have the matrix $V$ defined by (\ref{v}),
$$
V=\left( \ba{cccccccc} (-d_1+d_8)/2& (d_1+d_8)/2 & 0& 0& 0& 0& 0&
0\\(d_1+d_8)/2& (-d_1+d_8)/2& 0& 0& 0& 0& 0&0\\0& 0& 0& d_2& 0& 0&
0& 0\\0& 0& d_7& 0& 0& 0& 0& 0\\0& 0& 0& 0& 0& d_3& 0& 0\\0& 0& 0&
0& d_6& 0& 0& 0\\0& 0& 0& 0& 0& 0& 0& d_4\\0& 0& 0&0& 0& 0& d_5& 0
\ea\right).
$$
Therefore
$$
\tilde{V}_{1|23}=\left( \ba{cccccccccccccccc}
\frac{-d_1+d_8}{2}& \frac{d_1+d_8}{2}& 0& 0& \frac{d_1+d_8}{2}&
\frac{-d_1+d_8}{2}& 0& 0& 0& 0& 0& d_2& 0& 0& d_7& 0\\
0& 0& 0& 0& 0& 0& 0& 0& 0& 0& 0& 0& 0& 0& 0& 0\\
0& 0& 0& 0& 0& 0& 0& 0& 0& 0& 0& 0& 0& 0& 0& 0\\
0& d_3& 0& 0& d_6& 0& 0& 0& 0& 0& 0& d_4& 0& 0& d_5& 0 \ea\right),
$$
$$
\tilde{V}_{12|3}=\left( \ba{cccccccccccccccc}
(-d_1+d_8)/2& (d_1+d_8)/2& (d_1+d_8)/2& (-d_1+d_8)/2\\0& 0& 0& 0\\
0& 0& 0& 0\\0& 0& 0& 0\\0& 0& 0& 0\\
0& d_2& d_7& 0\\0& 0& 0& 0\\0& 0& 0& 0\\0& 0& 0& 0\\0& 0& 0& 0\\
0& d_3& d_6& 0\\0& 0& 0& 0\\0& 0& 0& 0\\0& 0& 0& 0\\0& 0& 0& 0\\
0& d_4& d_5& 0 \ea\right),
$$
obviously, when $d_1=-d_2=d_3=-d_4=-d_5=d_6=-d_7=d_8=1,
r(\tilde{V}_{1|23})=r(\tilde{V}_{12|3})=1$. Hence $\rho$ and
$\rho^\prime$ are  equivalent under local unitary transformations.

Now we consider the multipartite case.
A general pure state on $H_1\otimes H_2\otimes\dots\otimes H_M$ is
of the form
\begin{equation}
\label{pm} |\Psi_M\ra=\sum_{k=1}^{M}\sum_{{i_k}=1}^{N_k} a_{i_1
i_2 \dots i_M}|e_{i_1}\ra\otimes |f_{i_2}\ra\otimes\dots\otimes
|g_{i_M}\ra, \hspace{0.6cm}a_{i_1 i_2\dots i_M}\in {\Bbb C}
\end{equation}
with $\sum a_{i_1 i_2\dots i_M}a_{i_1 i_2 \dots i_M}^\ast=1$,
$|e_{i_1}\ra,~
|f_{i_2}\ra,~...,~|g_{i_M}\ra$, $i_k=1,2,...,N_k$, $k=1,2,...,M$,
the corresponding orthonormal basis
of complex Hilbert spaces $H_1,~H_2,~...,~H_M$.
Two density matrices $\rho$ and $\rho^\prime$ are said to be
equivalent under local unitary transformations if there exist
unitary operators $U_1$ on $H_1$, $U_2$ on $H_2$, $\cdots$, and
$U_M$ on $H_M$ such that
$\rho^\prime=(U_1\otimes U_2\otimes\cdots\otimes U_M)\rho(U_1\otimes
U_2\otimes\cdots\otimes U_M)^\dag$.
For any non-degenerate density matrices we have

{\sf [Theorem 2].} Let $\rho$ and $\rho^\prime$ be two non-degenerate density
matrices on $H_1\otimes H_2\otimes\dots\otimes H_M$ and $X$ and $Y$ the unitary matrices that
diagonalize $\rho$ and $\rho^\prime$ respectively,
$\rho=X\Lambda X^\dag$, $\rho^\prime=Y\Lambda Y^\dag$, where
$\Lambda=diag(\lambda_1,\lambda_2,...,\lambda_{N_1N_2\cdots N_M})$.
$\rho$ and $\rho^\prime$ are equivalent under local unitary transformations
if and only if the $N_1N_2\cdots N_M \times N_1N_2\cdots
N_M$ unitary matrix $V=XDY^\dag$, $D=diag(e^{i\theta_1},\,e^{i\theta_2}
,...,e^{i\theta_{N_1N_2\cdots N_M}})$, satisfies
$$
r(\tilde{V}_{1|2\cdots
M})=r(\tilde{V}_{12|3\cdots M})=
\cdots=r(\tilde{V}_{12\cdots {M-1}|M})=1,
$$
where
$$
\ba{l}
(\tilde{V}_{1|2\cdots M})_{i_1i_1',i_2\cdots i_Mi_2'\cdots
i_M'}=(V)_{i_1i_2\cdots i_M,i_1'i_2'\cdots i_M'},\\[2mm]
(\tilde{V}_{12|3\cdots M})_{i_1i_2i_1'i_2',i_3\cdots
i_Mi_3'\cdots i_M'} =(V)_{i_1i_2\cdots i_M,i_1'i_2'\cdots i_M'},\\[3mm]
\qquad \qquad \cdots \cdots \cdots  \cdots\\[1mm]
(\tilde{V}_{12\cdots {M-1}|M})_{i_1i_2\cdots
i_{M-1}i_1'i_2'\cdots i_{M-1}',i_Mi_M'}=(V)_{i_1i_2\cdots i_M,i_1'i_2'\cdots
i_M'}.\ea
$$

We have studied the equivalence of multipartite quantum mixed states under
local unitary transformations in terms of
analysis of fixed point subgroup and tensor
decomposability of certain matrices. A criterion for the
equivalence of non-degenerate mixed multipartite quantum mixed states under
local unitary transformations has been presented.
In fact this approach works for general multipartite mixed states.
But then in stead of the rank $r(\tilde{V})$, one has to verify if $r(\tilde{V_0})$
could be one for all possible matrices $A_{n_i}$, which is again
complicated. The problem is dramatically simplified
when the degeneracy of the related density matrices is
reduced. In particular, for the non-degenerate case, two density matrices are
easily verified whether they are equivalent or not under
local unitary transformation.

\vspace{1.0truecm}

\noindent {\bf Acknowledgments} The first and last named author gratefully
acknowledges the support provided by
the China-Germany Cooperation Project 446 CHV 113/231, ``Quantum
information and related mathematical problems".

\end{document}